\documentclass[12pt,preprint]{aastex}

\newcommand{\myemail}{jricardo.rizzo@uem.es}

\newcommand{\kms}{km s$^{-1}$}

\newcommand{\htwo}{H$_2$}

\newcommand{\tk}{$T_{\mathrm K}$}

\newcommand{\vlsr}{$V_{\rm LSR}$}

\newcommand{\cthd}{C$_3$H$_2$}
\newcommand{\cdh}{C$_2$H}

\slugcomment{Accepted by \apj}

\shorttitle{Chemistry of Mon R2}
\shortauthors{Rizzo et al.}

\begin{document}

%% LaTeX will automatically break titles if they run longer than
%% one line. However, you may use \\ to force a line break if
%% you desire.

\title{Insights into the carbon chemistry of Mon R2}

%% Use \author, \affil, and the \and command to format
%% author and affiliation information.
%% Note that \email has replaced the old \authoremail command
%% from AASTeX v4.0. You can use \email to mark an email address
%% anywhere in the paper, not just in the front matter.
%% As in the title, use \\ to force line breaks.

\author{J. R. Rizzo}
\affil{Departamento de F\'{\i}sica, Universidad Europea de Madrid, 
Urb.~El Bosque, Tajo s/n, E-28670 Villaviciosa de Od\'on, Spain}
\email{\myemail}

\author{A. Fuente and S. Garc\'{\i}a-Burillo}
\affil{Observatorio Astron\'omico Nacional, Aptdo.~Correos 112, E-28803
Alcal\'a de Henares, Spain}

%% Mark off your abstract in the ``abstract'' environment. In the manuscript
%% style, abstract will output a Received/Accepted line after the
%% title and affiliation information. No date will appear since the author
%% does not have this information. The dates will be filled in by the
%% editorial office after submission.

\begin{abstract}
Aiming to learn about the chemistry of the dense PDR around the ultracompact (UC)
\ion{H}{2} region in \objectname{Mon R2}, we have observed a series of mm-wavelength 
transitions of \cthd\ and C$_2$H. In addition, we have traced the distribution of 
other molecules, such as H$^{13}$CO$^{+}$, SiO, HCO, and HC$_3$N. These data, 
together with the reactive ions recently detected, have been considered to 
determine the physical conditions and to model the PDR chemistry. We then 
identified two kind of molecules. The first group, formed by the reactive ions 
(CO$^+$, HOC$^+$) and small hydrocarbons (C$_2$H, C$_3$H$_2$), traces the surface 
layers of the PDR and is presumably exposed to a high UV field (hence we called it 
as ``{\it high UV}'', or HUV). HUV species is expected to dominate for visual 
absorptions $2<A_{\mathrm V}<5$\,mag. A second group (less exposed to the UV 
field, and hence called ``{\it low UV}'', or LUV) includes HCO and SiO, and is 
mainly present at the edges of the PDR ($A_{\mathrm V}>5$\,mag). While the 
abundances of the HUV molecules can be explained by gas phase models, this is not
the case for the studied LUV ones. Although some efficient gas-phase reactions might
be lacking, grain chemistry sounds like a probable mechanism able to explain the
observed enhancement of HCO and SiO. Within this scenario, the interaction of 
UV photons with 
grains produces an important effect on the molecular gas chemistry and constitutes 
the first evidence of an ionization front created by the UC \ion{H}{2} region 
carving its host molecular cloud. The physical conditions and kinematics of the 
gas layer which surrounds the UC \ion{H}{2} region were derived from the HUV 
molecules. Molecular hydrogen densities $>4\,10^6$ cm$^{-3}$ are required to
reproduce the observations. Such high densities suggest that the \ion{H}{2} 
region could be pressure-confined by the surrounding high density molecular gas.
\end{abstract}

%% Keywords should appear after the \end{abstract} command. The uncommented
%% example has been keyed in ApJ style. See the instructions to authors
%% for the journal to which you are submitting your paper to determine
%% what keyword punctuation is appropriate.

%% Authors who wish to have the most important objects in their paper
%% linked in the electronic edition to a data center may do so in the
%% subject header.  Objects should be in the appropriate "individual"
%% headers (e.g. quasars: individual, stars: individual, etc.) with the
%% additional provision that the total number of headers, including each
%% individual object, not exceed six.  The \objectname{} macro, and its
%% alias \object{}, is used to mark each object.  The macro takes the object
%% name as its primary argument.  This name will appear in the paper
%% and serve as the link's anchor in the electronic edition if the name
%% is recognized by the data centers.  The macro also takes an optional
%% argument in parentheses in cases where the data center identification
%% differs from what is to be printed in the paper.


%% From the front matter, we move on to the body of the paper.
%% In the first two sections, notice the use of the natbib \citep
%% and \citet commands to identify citations.  The citations are
%% tied to the reference list via symbolic KEYs. The KEY corresponds
%% to the KEY in the \bibitem in the reference list below. We have
%% chosen the first three characters of the first author's name plus
%% the last two numeral of the year of publication as our KEY for
%% each reference.

\section{Introduction}
\objectname{Mon R2} is a nearby \citep[$D=830$\,pc;][]{her76}
complex star forming region. It hosts an ultracompact (UC) {\sc Hii} region 
near its center, powered by the infrared source \objectname{Mon R2 IRS\,1}.
The molecular content of this region has been the subject of several observational
studies in the last decade. So far, it is known a huge CO bipolar outflow 
\citep{mey91}, $\sim 15'$ long (=\,3.6\,pc), which is probably a relic
of the formation of the B0V star associated to IRS1 \citep{mas85,hen92}.

\objectname{Mon R2} is an excellent laboratory to learn about the physical conditions 
and kinematics of an extreme PDR \citep[G$_0$ = 5 10$^5$ in units of Habing 
field,][]{riz03}. Continuum observations at 6cm, reported by 
\citet{woo89}, show that the UC {\sc Hii} region is 
highly asymmetric, has a cometary shape and reaches its maximum 
toward its exciting star, \objectname{Mon R2 IRS\,1}. The CS 7$\rightarrow$6 emission map 
from \citet{cho00} shows that the UC {\sc Hii} region is 
located inside a cavity and bound by a dense molecular ridge (see Fig.~1). 
This picture is consistent with infrared images of the region, which 
show an almost spherical distribution of hot dust surrounded by the 
molecular cloud.

The physical and chemical conditions of the PDRs associated to  UC {\sc Hii}
regions remain not enough known, mainly due to important observational 
problems: (a) These PDRs are too thin to be resolved by single-dish 
telescopes; (b) They are usually located in very complex
star-forming regions, where confusion with the parent molecular cloud, hot cores, 
and outflows may be considerable. The study requires 
specific PDR tracers which allow us to separate their emission from
other sources.

Reactive ions and small hydrocarbons have been proposed as abundant species, 
which may trace the hot ionized/molecular gas interface \citep{ste95}. The 
detection of the reactive ions CO$^+$ and HOC$^+$ is 
almost unambiguously associated to regions with a high ionizing
flux, either PDRs or XDRs \citep{fue03,riz03,use04,sav04}.
In the case of \objectname{Mon R2}, \citet{riz03} have recently reported 
the detection of the reactive ions CO$^+$ and HOC$^+$ towards the
peak of the UC {\sc Hii} region. In addition to chemical
arguments, both the morphology and velocity profile of the lines suggested
a link between the {\sc Hii} region and these species. Both ions were
detected towards the peak of the UC {\sc Hii} regions and remain undetected in
the molecular cloud. Furthermore, the velocity profile of these lines is different
from those of other dense gas tracers. The reactive ions have an intense
emission peak at 10.5 km s$^{-1}$ which is not detected in the other molecules 
observed. Therefore, the emission of this velocity component may
arise in narrow layers of molecular gas surrounding the {\sc Hii} region, where 
the chemistry is dominated by UV photons. 

Small hydrocarbons have enhanced abundances in PDRs \citep{fue03,fue05,tey04,pet05}.
In particular, \citet{tey04} have detected c-C$_3$H$_2$ and C$_4$H towards 
the Horsehead nebula with an abundance one order of magnitude larger than that 
predicted by gas-phase PDR models. This is clearly seen by comparing the 
[c-C$_3$H$_2$]/[HC$_3$N] ratio in PDRs and dark clouds. While both species have 
similar abundances in dark clouds, the [c-C$_3$H$_2$]/[HC$_3$N] ratio is above 10 
in PDRs. Since both molecules are easily destroyed by photodissociation, this 
suggests the existence of an additional c-C$_3$H$_2$ formation mechanism in PDRs.

Enhanced abundances of some other compounds have also been related to the presence 
of intense UV fields into molecular clouds. \citet{sch88} measured a value 
of [HCO]/[H$^{13}$CO$^+$]\,=\,9.7 in the \ion{H}{2} region NGC\,2024. However, the same
authors report values significantly lower than 1 --i.e., an order magnitude below 
those found in NGC\,2024-- in Galactic clouds without developed \ion{H}{2} regions, or 
having no indication of star formation. More recently, \citet{sch01} have 
searched for HCO in a reduced sample of prototypical PDRs; the estimated 
[HCO]/[H$^{13}$CO$^+$] abundance ratios range from 30 (in the Orion bar) to 3 (in 
NGC\,7023). The largest HCO abundances are found in the Orion bar, the paradigm of 
interaction between an \ion{H}{2} region (M42) and its parent molecular cloud. 
\citet{gar02} obtained a high angular resolution 
image showing widespread HCO emission in the external galaxy M82. The enhanced HCO 
abundance ([HCO]/[H$^{13}$CO$^+$]\,$\sim$\,3.6) measured across the whole M82 disk 
was also interpreted in terms of a giant PDR of 650 pc size. 
 
SiO is known to be a privileged tracer of large-scale shocks in the interstellar 
medium \citep{mar97,gar00,gar01}. Its fractional abundance is enhanced by more 
than 3 orders of magnitude in shocked regions relative to its abundance in 
quiescent clouds where it takes values $\leq$ 10$^{-12}$. \citet{sch01} observed 
SiO towards a sample of PDRs and obtain fractional abundances of $\sim$10$^{-11}$ 
in these regions, i.e., a factor of 10 larger than that in dark clouds.

In this paper, we present observations of a selected set of molecular species 
(C$_2$H, c-C$_3$H$_2$, HC$_3$N, H$^{13}$CO$^+$, HCO, and SiO) which are thought 
to be good probes of the different layers of the molecular gas in PDRs. In this 
way, we have performed a study of the physical conditions, the chemistry, and the 
kinematics of the molecular gas surrounding the UC {\sc hii} region.

%% ================= O B S E R V A T I O N S ==============================

\section{Observations}

The observations were carried out using the IRAM 30m radio telescope in Pico de 
Veleta (Spain) during July 2002 and August 2003. The multi-receiver capability 
of the telescope was used for these observations. The observed rotational 
transitions, frequencies, beam sizes, and main-beam efficiencies are shown in 
Table 1. The backends used for the observations were an autocorrelator split in 
several parts, a 256 $\times$ 100 kHz and two 512 $\times$ 1 MHz filter-banks. 
All the lines but H$^{13}$CO$^+$ 1$\rightarrow$0 and \cthd\ 5(1,4)$\rightarrow$4(2,3)
have been measured with two spectral resolutions, namely the lower spectral 
resolution provided by the 1 MHz filter-bank and the higher one, 
78 kHz--100 kHz, provided by the autocorrelator and the 100 kHz filter bank. 
The H$^{13}$CO$^+$ 1$\rightarrow$0 and \cthd\ 5(1,4)$\rightarrow$4(2,3) lines
have only been measured with the 1 MHz filter-bank, providing the lower
spectral resolution. The \cthd\ 6(1,6) $\rightarrow$5(0,5) [ortho] and 
6(0,6)$\rightarrow$5(1,5) [para] are blended.

The observational strategy was as follows. We have mapped the UC \ion{H}{2} 
region \objectname{Mon R2} and its surroundings in millimeter transitions of H$^{13}$CO$^+$, 
HC$_3$N, HCO, and SiO. The C$_2$H and \cthd\ transitions were observed along 
a strip transversal to the UC \ion{H}{2} region, at an inclination angle (east to 
north) of 135\degr, which is approximately the symmetry axis of \objectname{Mon R2}. The 
mapped area ($72\arcsec\times72\arcsec$) and strip positions are sketched in 
Fig.~1, overlaid on the CS 7$\rightarrow$6 map of \citet{cho00} and
the 6 cm-continuum map of \citet{woo89}. The relative offsets are
referred to \objectname{Mon R2 IRS1}, 
(RA, Dec.)$_{2000}$=($06^{\rm h}07^{\rm m}46\fs2,-06\degr23\arcmin08\farcs3$).
Longer integration time was devoted to observe two particular positions: the
maximum of the continuum emission, at (0\arcsec, 0\arcsec), and the maximum
of the molecular envelope, at (+10\arcsec, -10\arcsec). These two positions 
were also observed in the reactive ions CO$^+$ and HOC$^+$, and already 
published in a recent paper \citep{riz03}.

%% ========= R E S U L T S ==============================================

\section{Results}
\subsection{\cthd\ and \cdh\ along the strip}

The small hydrocarbons \cdh\ and \cthd\ have been observed along a strip crossing 
the ionization bar in \objectname{Mon R2}. After the hyperfine-structure analysis --which
compares the relative intensities of each component--, we have found that the
\cdh\ triplet emission at 87.3 GHz is optically thin at all positions. Based on this
result, we will deal in the following just with the most intense hyperfine 
component ($F=2-1$), referring to this as ``the \cdh\ 1$\rightarrow$0 line''. Some
of the \cdh\ and \cthd\ observed spectra are shown in Fig.~2. The Gaussian fits 
to a single component are also shown in Table 2 for all the lines. The 
C$_2$H 1$\rightarrow$0 line shows emission at high velocities (\vlsr~$<8$ km\,s$^{-1}$ 
and \vlsr~$>12$ km\,s$^{-1}$) that is not detected (or only marginally detected) in 
the C$_3$H$_2$ lines. These velocity ranges coincide with those of the red-shifted 
and blue-shifted emission in the large-scale, now inactive, outflow associated 
with IRS 1 \citep{mey91,gia97}. We have not detected emission at high velocity in 
any other of the PDR tracers studied in this paper and \citet{riz03}. Thus, we 
consider that the high velocity C$_2$H emission is not related to the UC 
\ion{H}{2} region but to the low density gas that share the kinematics of the 
large-scale outflow. Since we are mainly interested in the study of the PDR 
surrounding the UC \ion{H}{2} region, we have only used the C$_2$H emission in the 
range 8--12 km\,s$^{-1}$ in our calculations.
 
Within the considered velocity range, the velocities and spatial distribution 
of the C$_2$H 1$\rightarrow$0 and C$_3$H$_2$ 2$\rightarrow$1 lines are 
remarkable similar, which suggest that both lines trace essentially the same 
region. However, the emission of the C$_3$H$_2$ 2$\rightarrow$1 and the 
C$_3$H$_2$ 6$\rightarrow$5 lines have different kinematics and spatial 
distribution, which became eloquent when looking at the central velocities in 
Table 2 and the shaded areas in Fig.~2. The C$_3$H$_2$ 2$\rightarrow$1 line 
is narrower than the 6$\rightarrow$5 line along the whole strip. Furthermore, 
the central velocity of the 2$\rightarrow$1 line is lower than that of the 
6$\rightarrow$5 line towards the UC \ion{H}{2} region. When looking at the 
areas, it is remarkable the different spatial distribution of the 
2$\rightarrow$1 and 6$\rightarrow$5 lines. The 2$\rightarrow$1 emission (a low 
excitation line) is clearly lower at the central part of the strip, while the 
6$\rightarrow$5 emission (a high excitation line) is particularly intense there. 

This behavior is better explained by assuming the existence of {\it at least} 
two gas components characterized by different velocities. The blue velocity 
component (hereafter referred to as C1), emitting at velocities from $\approx$ 
8 to 10 \kms, is dominated by the emission of the 2$\rightarrow$1 line along the 
whole strip, but it is not particularly intense in the 6$\rightarrow$5 line. 
This component remains quite uniform along the strip. The red velocity component, 
with emission from $\approx$ 10 to 12 \kms and hereafter referred to as C2, has  
intense emission in the 6$\rightarrow$5 line but it is very weak in the 
2$\rightarrow$1 line. C2 is especially intense towards the inner \ion{H}{2} region.
The distribution of both components can be interpreted as C1 associated with 
the foreground molecular cloud and C2 associated with the \ion{H}{2} region, or the 
interface between it and the molecular cloud.

This trend is illustrated in Fig.~3, where we plot the integrated intensity of the
\cthd\ 6$\rightarrow$5 line (Fig.~3a) and the 
6$\rightarrow$5/2$\rightarrow$1 line intensity ratio (hereafter referred to as
R$_{62}$) for each component (Fig.~3b). R$_{62}$ remains quite constant and
$\sim$ 1 for C1 along the strip. However, it has significant
variations in C2. The values of R$_{62}$ are $\sim 0.6$ outside the 
UC \ion{H}{2} region and $>4$ towards the inner region. The Fig.~3b clearly separates both 
components, and shows that C2 is tracing highly excited molecular gas, linked to the 
UC \ion{H}{2} region. This splitting of the values of R$_{62}$ is also in agreement with 
the spatial distribution of both components.

Of course, the different beams involved in both lines affect the computation of
R$_{62}$. 
%However, it should be noted that the high values of R$_{62}$ in C2 
%remain over three positions, separated by $\sim 15''$ each. In other words, these 
%particular physical conditions are present over $\sim 30''$, which is larger than 
%the beam of the \cthd\ 6$\rightarrow$5 line. 
We could estimate the beam dilution 
by convolving the \cthd\ 6$\rightarrow$5 emission of the C2 down to the angular 
resolution of the 2$\rightarrow$1 line. This one-dimensional smoothing was 
performed along the symmetry axis, which shows the largest variations in other 
molecular tracers. The results are shown by dashed lines in Fig.~3. As expected, 
the distribution of the 6$\rightarrow$5 line intensity peaks in the southeastern 
border of the UC \ion{H}{2} region. Furthermore, it is remarkable that R$_{62}$ remains 
$> 5$ in the three central positions. In other words, the high value of R$_{62}$ 
there, even smoothed, strengthens the mutual relationship between C2 and the UC 
\ion{H}{2} region. 

\subsection{Maps}
 
We have mapped the region indicated in Fig.~1 in several rotational lines.  
Fig.~4 shows the resultant maps in H$^{13}$CO$^+$ 1$\rightarrow$0 (Fig.~4a), 
SiO 2$\rightarrow$1(Fig.~4b), HCO 1$\rightarrow$0 (Fig.~4c), and 
HC$_3$N 10$\rightarrow$9 (Fig.~4d). Superimposed are 
the 6 cm-continuum emission from \citet{woo89}.
All the maps were constructed by integrating the lines along the total 
velocity range of emission.
We note remarkable differences in the distribution of the different species.
The H$^{13}$CO$^+$ and the HC$_3$N emission clearly follow the high density 
gas, properly traced by the CS 7$\rightarrow$6 emission (see Fig.~1) and  
reach a minimum toward the center of the UC \ion{H}{2} region. This is consistent with 
the interpretation of H$^{13}$CO$^+$ and HC$_3$N as being photodissociated 
within the UC \ion{H}{2} region.

However, the other molecules present different behaviors. The SiO and HCO 
emissions seem to be located preferably at intermediate positions between the 
molecular envelope and the \ion{H}{2} region. Although this effect should be
regarded as tentative, because the spatial difference between the continuum
peak and the SiO distribution is about half a beam, the systematic difference
looks evident, and clearly the SiO or HCO maps are different to those of 
H$^{13}$CO$^+$ or HC$_3$N. This result is in line with 
\citet{sch01}, who also found moderately enhanced SiO and HCO emissions close to 
the ionization front in the prototypical \ion{H}{2} region M42. However, the two 
species peak at different positions. The HCO maxima appears almost exactly at the 
SiO minimum, and both emissions completely bound the inner UC \ion{H}{2} region.

In Fig.~5 we show all the spectra we have observed towards the (0\arcsec,0\arcsec) 
position \citep[this paper and][]{riz03}. The dashed vertical line traces the 
limits of C1 and C2. Unfortunately, some of the lines at (0\arcsec,0\arcsec) are
rather weak, which yields a low S/N ratio. Even though, a quick look suggests 
the existence of two different behaviors in the shape of the observed lines. 
A first group of lines has the maximum intensity at velocities $<$ 10 \kms, close
to the C1 velocity; this is the case for the SiO and HCO lines. A second group 
peaks at velocities $>10$ km s$^{-1}$ and are especially 
intense in C2; the C$_3$H$_2$ 6$\rightarrow$5, CO$^+$ 2$\rightarrow$1 and 
HOC$^+$ 1$\rightarrow$0 belongs to this group. The strong difference between
the SiO lines at C2 velocities should be more carefully analyzed in future 
observations.

After a joint view of the maps (Fig.~4) and the spectra (Fig.~5), we see that 
the SiO and HCO emitting gas do not seem to share
the same volume with the reactive ions and small hydrocarbons, but to arise in 
a more shielded layer of the PDR, which is preferably at the velocity of C1. As 
further discussed in Sects.~5 and 6, different excitation conditions and 
chemistry are associated to this shielded layer and C2.

\section{Physical conditions: LVG results for the hydrocarbons}

We have developed a LVG code to estimate the main physical parameters of the 
\cthd\ molecule. The Einstein A- coefficients have been taken from 
\citet{cha03}. The collisional rates are from \citet{cha00}, who provided
values for kinetic temperatures (\tk) of 30, 60, 90, and 120\,K.
Our LVG code treats the ortho- and para-C$_3$H$_2$ as 
different species. In most positions, we have detected only the 
2(1,2)$\rightarrow$1(0,1) [ortho] and the 6$\rightarrow$5 (a blending of the 
6(1,6)$\rightarrow$5(0,5) [ortho] and 6(0,6)$\rightarrow$5(1,5) [para]) lines. 
Then, we need to assume a value of the C$_3$H$_2$ ortho-to-para ratio (OTPR) in 
our calculations. In order to have an estimate of the OTPR, we have used the 
6$\rightarrow$5/5$\rightarrow$4 line intensity ratio (hereafter R$_{65}$) 
towards the (0$"$,0$"$) and ($+10"$,$-10"$) positions. The observed value,
R$_{65}\sim 1.4\pm 0.3$, is compatible with R$_{65}=4/3$, which is expected for 
the standard value of 3. Therefore, we have assumed OTPR\,=\,3 in our LVG 
calculations hereafter.

The LVG code was run for different sets of \tk, from 10 to 150\,K. For \tk\
different to those tabulated by \citet{cha00}, we have interpolated or 
extrapolated between the two closest temperatures.
A minimum value of 30\, K for \tk\ is needed to reproduce the 
6$\rightarrow$5/2$\rightarrow$1 line ratio (hereafter referred to as R$_{62}$) 
both in C1 and C2. For \tk\,$>50$\,K, R$_{62}$ is weakly dependent on the assumed
kinetic temperature and traces mainly the hydrogen density. In Table 3, we show 
the LVG results assuming \tk\,$=50$\,K, the kinetic temperature derived by 
\citet{gia97} from multiline molecular observations. 

We have obtained rather uniform values of both the molecular density ($n$(\htwo))
and the \cthd\ column density ($N$(\cthd)) in  C1 across the observed strip. The 
derived $n$(\htwo) is around 5\,$10^5$ cm$^{-3}$, in agreement with 
results derived from other tracers \citep{cho00,riz03}, and the $N$(\cthd) are in 
the range 1--3\,$10^{12}$ cm$^{-2}$. A different behavior is observed in C2.
The C2 hydrogen density increases from a few 10$^5$ cm$^{-3}$ outside the UC 
\ion{H}{2} region to $>4\,10^6$ cm$^{-3}$ in the three central positions. This limit 
is almost insensitive to \tk\ in the range 50 to 150\,K. Thus, the larger values 
of the R$_{62}$ observed towards the \ion{H}{2} region cannot be due to a higher 
kinetic temperature. The Fig.~6 shows the LVG results from two of the runs, 
corresponding to values of \tk\ of 60 and 120\,K. The shadowed areas indicate 
the range of observed values towards the ($0"$,$0"$) position. While $N$(C$_3$H$_2$) 
is similar in C1 and C2, the C2 density is a factor $>5$ greater than in C1.

Recent results of \citet{jaf03} and near-IR images show a shell-like structure, 
intense in mid- and near-IR, of a few arcsec width, wrapping up the UC 
\ion{H}{2} region. It is indeed possible that C2 is tracing the {\it high density 
molecular gas} associated to this feature. This would also explain the spatial 
distribution of $N$(C$_3$H$_2$) in C2 which slightly increases towards the border 
of the \ion{H}{2} region, the expected behavior of a molecular shell surrounding the 
ionized gas.

We have estimated the C$_2$H column density ($N$(\cdh)) across the studied 
strip using the LVG code and assuming $n$(\htwo)\,=\,10$^6$ cm$^{-3}$ and 
\tk\,=\,50\,K. We have adopted the cross sections provided by \citet{gre74} for
the HCN molecule, and then used the IOS (Infinite-Order Sudden) approximation
for molecular collision dynamics to get the collisional coefficients between 
different levels. A similar procedure was followed by \citet{tru87}, who also
provided the Einstein A- coefficients for \cdh.

The C1 and C2 components have been separated towards
the (0$"$,0$"$) and ($+10"$,$-10"$). Most of the \cdh\ emission seem to be 
associated to C1 outwards these positions. The results are shown in Tables 3 and 
4. $N$(C$_2$H) varies from 2 to 4 10$^{14}$ cm$^{-2}$, decreasing toward the 
center of the \ion{H}{2} region, as expected in a region where the molecules are 
photodestroyed by a high incident UV flux. When compared to C$_3$H$_2$, the
[\cthd]/[\cdh] ratio slightly increases towards the center; an exception is
the last point of the strip, which has a considerable lower S/N ratio (see 
Table 3). Comparing both components, the [\cthd]/[\cdh] ratio is three times 
greater in C2 than in C1 (see Table 4). As we will refer in Sect.~ 6.1, the 
significantly larger [\cthd]/[\cdh] ratio in C2 reveals that the chemistry of 
the molecular gas is heavily affected by the UV radiation. 

\section{H$^{13}$CO$^+$, SiO and HC$_3$N fractional abundances}

The H$^{13}$CO$^+$, SiO and HC$_3$N column densities have been estimated
using an LVG code and assuming \tk\,=\,50\,K and $n$(H$_2$)\,=\,10$^6$\,cm$^{-3}$.
Under these physical conditions, the excitation temperature of the H$^{13}$CO$^+$
1$\rightarrow$0 line is $\sim$\,20\,K. We have also derived the HCO column density 
by assuming optically thin emission, LTE conditions and a rotation temperature of 
20\,K. The uncertainty in the excitation temperature of HCO should not affect the 
fractional abundances of these molecules. The results for some selected positions 
are shown in Table 5. 

Significant changes in the fractional abundances are detected across the UC 
\ion{H}{2} region. In particular, the HC$_3$N abundance decreases by a factor of 
$\sim$3 towards the UC \ion{H}{2} region (X(HC$_3$N$)\sim$ a few 10$^{-10}$), when 
compared towards the dense molecular cloud (where X(HC$_3$N) $\sim$ 10$^{-9}$). 
This decrease of the HC$_3$N abundance in regions of enhanced UV field has also 
been found in other regions like the prototypical PDR of the Orion Bar 
\citep{rod98}, or the starburst galaxy M82 \citep{fue05}, and it is easily 
explained by photodissociation.

SiO has a different behavior. The SiO abundance is low towards the center of the 
UC \ion{H}{2} region (X(SiO)\,$\sim 5$\,10$^{-12}$). It is also low towards the dense 
molecular cloud, where we derive X(SiO)\,$<7$\,10$^{-12}$, in agreement with the 
values of the SiO abundance measured in dark clouds. However, it is one order of 
magnitude higher at intermediate positions between the UC \ion{H}{2} region and the 
molecular cloud; here, the SiO abundance is comparable to that found by 
\citet{sch01} towards the PDRs associated with the Orion Bar and S140. 

Similarly to SiO, the HCO abundance is also enhanced at the border of the UC \ion{H}{2} 
region. In particular the [HCO]/[H$^{13}$CO$^+$] ratio is 10 times larger in the 
position (0\arcsec,24\arcsec) than in the center of the UC \ion{H}{2} region and in the dense 
molecular cloud. We have measured a HCO abundance as high as $\sim10^{-9}$ towards 
the HCO peak. Such large values of the HCO abundance have been found 
at the border of the \ion{H}{2} region  NGC 2024 by \citet{sch88} and
in the Orion Bar by \citet{sch01}, suggesting that this molecule is specially 
abundant in the PDRs formed at the edges of the \ion{H}{2} regions. \citet{gar02} 
found large values of the HCO abundance in the starburst galaxy M82, where a 
low-density ionized component is filling a substantial fraction of its volume 
\citep[see also][]{sea96}.

\section{Photon-dominated chemistry in Mon R2}
\subsection{The surface layers of the PDR:  HUV species}

Our data show the existence of a high-density layer of gas associated to the 
UC \ion{H}{2} region in \objectname{Mon R2}. This layer is traced by C2 and 
seems to have different physical conditions from that of C1 and the molecular 
cloud. Furthermore, a different chemistry may also carry out in the layer 
traced by C2. In order to explore this idea, we have computed the column 
densities and fractional abundances for the two components, towards the 
positions (+10\arcsec,-10\arcsec) and (0\arcsec,0\arcsec), and for the molecules 
studied in this paper and \citet{riz03}. The results are shown in Table 4. 
Significant differences exist between the [CO$^+$]/[HCO$^+$], 
[HCO$^+$]/[HOC$^+$] and [C$_3$H$_2$]/[C$_2$H] ratios between C1 and C2. In order 
to characterize the chemistry of the two components, we have modeled the PDR 
using the plane-parallel model developed by Le Bourlot and collaborators 
\citep{leb93} with an UV field G$_0=5\,10^5$ in units of the Habing field and a 
total density ({\it i.e.}~atomic plus molecular) of 2\,10$^6$ cm$^{-3}$ 
\citep[][~and this paper]{riz03}. The main results are sketched in Fig.~7.

The molecules CO$^+$, HOC$^+$, C$_2$H, and C$_3$H$_2$ are known to be good 
tracers of the surface layers ($A_{\mathrm V}<5$\,mag) of the PDRs 
\citep[see e.g.~][]{ste95}. These 
environments are highly exposed to an intense UV field, and hence we will refer 
to these molecules as ``{\it high UV}'' species (HUV for briefing). In 
particular, the reactive ion CO$^+$ is only expected to reach significant 
abundances at a visual extinction, $A_{\mathrm V}<2$\,mag \citep{fue03,riz03}. 
In Fig.~7a we show the CO$^+$ and HCO$^+$ fractional abundances as a function of the 
visual extinction for our model. As expected, the abundance of CO$^+$ is only 
significant at $A_{\mathrm V}<2$\,mag, and just at this low visual extinction, 
the predicted  [CO$^+$]/[HCO$^+$] ratio agrees with that measured in C2 towards 
the \ion{H}{2} region (light gray region in Fig.~7a). The [CO$^+$]/[HCO$^+$] 
ratio is more than one order of magnitude greater in C2 than in C1 (Table 4), 
which indicates that C1 arise in a more shielded layer of the molecular gas having 
$A_{\mathrm V}>4$\,mag (dark gray region in Fig.~7a).

The [HCO$^+$]/[HOC$^+$] ratio is a factor of 2 lower in C2 than in C1. Low values 
of the [HCO$^+$]/[HOC$^+$] ratio are associated to highly ionized regions 
\citep{use04,sav04,fue05}. Unfortunately, the reactive ion HOC$^+$ is not included 
in Le Bourlot's model, but we can estimate its abundance in an indirect way.
The [HCO$^+$]/[HOC$^+$] ratio is strongly dependent on the electron abundance. 
In order to have [HCO$^+$]/[HOC$^+$] $\sim 450$, it is required an electron 
abundance, X(e$^-$)\,$>2\,10^{-5}$ \citep{use04}. Using our model, we conclude 
that these high electron abundances are only reached at $A_{\mathrm V}<4$\,mag. 
Thus, the [CO$^+$]/HCO$^+$] and the [HCO$^+$]/[HOC$^+$] ratios support the 
interpretation of C2 as arising from a layer of molecular gas located at a visual 
extinction $<4$\,mag from the advancing ionization front. A new PDR model 
including reactions of HOC$^+$ is under development, and it will directly 
predict the [HCO$^+$]/[HOC$^+$] ratio, as a function of global physical conditions 
(UV field, density, extinction) in a near future.

The [C$_3$H$_2$]/[C$_2$H] ratio also changes between C1 and C2. In fact, this is 
a factor of 2--3 larger in C2 than in C1. Our chemical model predicts that the 
[C$_3$H$_2$]/[C$_2$H] ratio is maximum at a visual extinction between 2 and 4\,mag 
(see Fig.~7). These values of the visual extinction are in agreement with those 
derived from the measured [HCO$^+$]/[HOC$^+$] ratio. Thus, it seems reasonable 
that the emission of C2 arises in the same layer of dense gas than HOC$^+$ 
($A_{\mathrm V}<4$\,mag), while CO$^+$ arises in even a more external layer 
($A_{\mathrm V}<2$\,mag). However, our model fails to predict values of 
[C$_3$H$_2$]/[C$_2$H] as large as those measured in C2 even at this low visual 
extinction. The maximum value of the [C$_3$H$_2$]/[C$_2$H] ratio predicted by the 
model is $\sim$ 0.01, three times lower than those measured in C2. (see Fig.~7b 
and Table 4). A failure of PDR models to account for the observed C$_3$H$_2$ 
abundance has already been commented in previous works 
\citep{fue03,tey04,pet05}. The existence of an additional formation mechanism for 
C$_3$H$_2$ linked to the PAHs photodestruction has been proposed as an alternative 
to account the large C$_3$H$_2$ fractional abundances in PDRs.

Summarizing, we can infer from the comparison of our observations of HUV molecules
and model calculations that C2 is tracing the dense molecular gas located at a 
visual extinction $<4$\,mag from the ionization front, while C1 is mainly tracing 
the more shielded part ($A_{\mathrm V}>5$\,mag) of the PDR. Consequently, the UC \ion{H}{2} 
region is surrounding by a thin layer ($<4\ 10^{21}$ cm$^{-2}$) of dense 
($>4\,10^6$ cm$^{-3}$) molecular gas, whose chemistry is heavily affected by 
the UV stellar radiation and kinetically well represented by C2.

\subsection{The back layers of the PDR: LUV species}

In contrast to the HUV species, the column densities of HCO and SiO decrease 
towards the UC \ion{H}{2} region, as expected in this harsh environment. 
Furthermore, their abundances are larger in C1 than in C2 (see 
[HCO]/[H$^{13}$CO$^+$] and [SiO]/[H$^{13}$CO$^+$] in Table 4), and show an 
enhancement at its edges (see Table 5). 

Clearly, both HCO and SiO do not share a behavior similar to the HUV molecules;
moreover, they are not particularly abundant in the cold envelope. Both molecules
seem to be mainly present in the shielded parts of the PDR ($A_{\mathrm V}>5$\,mag), 
and would belong to a family of species which can survive in those environments, less
exposed to the UV field than the HUV molecules. We hereafter call such molecules 
as ``{\it low UV}'' (or LUV for briefing) species. The abundance of our LUV 
molecules are unusually high and could not be explained by gas-phase models
\citep[see, for example,][]{leu84,ste95}. 
Equally puzzling is the fact that both molecules peak at different positions.

As we have already stated before, both SiO and HCO have --near their maxima-- 
abundances comparable to other PDRs, such as \objectname{S140} and Orion Bar 
\citep{sch01}. Moreover, the [HCO]/[H$^{13}$CO$^+$] abundance ratio near the HCO 
peak is in the range 10--15, higher than in \objectname{NGC\,2024} \citep{sch88} 
and \objectname{M82} \citep{gar02}. This enhancement can be due to a few reasons.
Firstly, some relevant chemical reactions leading to formation of SiO and HCO 
might not be considered in gas-phase models. It was already analyzed by several 
authors \citep[see][]{wal99,sch01} and sounds rather unlikely.
Even though, we should not disregard the hypothesis, mainly due to a possible
lack of complete chemical networks and rate coefficients.

So far, the most widely accepted alternative explanation arises by considering 
the grain chemistry. As many authors pointed out 
\citep[see, for example, the review by][]{cas05}, the dust plays a key role in 
driving the chemistry and the energetics in PDRs. \citet{wal99} have modeled the
Si chemistry in PDRs, and found that most of Si is in solid form at extinctions
greater than 3 mag. This issue was observationally confirmed by \citet{fue00} in
\objectname{NGC\,7023}, where a special distribution and enhancement of SiO was
detected. \citet{sch01} have included photodesorption of Si in their 
PDR model. SiO would later be produced in gaseous phase by reactions with OH and
O$_2$. By doing so, those authors could account for the observed SiO abundance in 
\objectname{Orion Bar} and \objectname{S140}. Laboratory measurements of the rate 
coefficient of the reaction Si + O$_2$ $\rightarrow$ SiO + O \citep{lep01}, 
followed by detailed calculations of Si chemistry, have confirmed the results by 
\citet{sch01}.

In the case of HCO, the photodesorption of H$_2$CO --followed by its 
photodissociation in gaseous phase-- have been claimed as a mechanism capable to 
counterbalance the photodissociation of HCO against the UV photons \citep{sch01}.
The authors could get an enhancement of HCO (at $A_{\mathrm V}>5$ mag) but not to 
the observed levels. This may be due to the need of more reliable rates for the 
processes involved, but also to the reduced network used in the gas-phase 
formation of this low-abundance species. We have measured similar abundances of 
HCO and hence are in a similar situation; the UV field intensity, greater in 
\objectname{Mon R2} than in Orion Bar
\citep{riz03}, does not seem to affect the HCO abundance.

However, photodesorption may not efficiently work in some other environments.
\citet{ruf01} have modeled both gas- and dust-chemistry in quiescent cores, and 
shown that the abundance of some species with lower binding energies than Si do not 
significantly change when including photodesorption. The modeled quiescent
cores, however, have a $G_0/n$ ratio at least two orders of magnitude lower than in 
\objectname{Mon R2}, because the densities are two orders of magnitude lower, and the 
incident UV field takes interstellar values.
Even though, their results should be taken 
into account, and a complete model like the \citeauthor{ruf01}'s one, 
under physical conditions more typical of PDRs, will surely improve our knowledge 
about the particular chemistry of LUV species and PDRs in general. Anyway, if SiO 
and HCO are formed after photodesorption of Si and H$_2$CO, respectively, we would 
expect other atoms and molecules do so, and would be detectable. At this respect, 
an interesting test may be to carry out an observational campaign to search for and 
study other LUV species throughout the whole field. A good candidate to belong
to the LUV group in NH$_3$, a widely recognized species linked to grain chemistry, 
whose abundance close to the borders of \ion{H}{2} regions is enhanced \citep{lar03}.

The different spatial distributions of our LUV molecules is another interesting
issue. Both molecules almost complete a ``ring'' between the UC \ion{H}{2} region
and the molecular envelope. However, they peak at different locations throughout
the ring: while SiO arises in the densest part of the ionization front, HCO peaks 
in the tail of the cometary shape. Different local conditions should exist in 
order to explain the variations in relative abundances of SiO and HCO.
A rather simple explanation is that HCO and SiO have a different desorption 
yield; in this case one would expect a layered structure in the morphology of 
these molecules. Unfortunately, the low angular resolution of the HCO and 
SiO maps prevent us from concluding about that. A second possibility is that 
a low-velocity shock associated with the ionization front contributes to enhance 
the HCO and/or SiO abundances. In the case of hot cores, this process was already 
analyzed by \citet{vit01}. The low velocity shocks contribute to release Si and/or 
H$_2$CO from the grain surfaces, and the gas phase chemistry is modified by the 
advancing shock front. The radical HCO is rapidly destroyed by O, but in shocked 
regions the O abundance is reduced because it is converted to H$_2$O. Furthermore, 
SiO is efficiently formed by the reaction Si $+$ OH $\rightarrow$ SiO $+$ H, 
being OH the photodissociation product of H$_2$O. Thus, the HCO and SiO 
abundances are very sensitive to the O/OH/H$_2$O relative fractional abundances 
and can change locally because of low velocity shocks and small inhomogeneities 
in the incident UV field. Finally, a variation in the grain composition may 
account for the differences between the HCO-peak to the SiO-peak. 

Observations of dust emission may help to
solve this interesting puzzle in \objectname{Mon R2}. \citet{kra01} have
studied this region in the mid-IR, from 8.2 to 20.6\,$\mu$m. Besides the IR sources,
continuum dust emission is dominated by a ring-like structure which roughly bounds
the ionized region. While the dust temperature is rather uniform in the area 
(100--115\,K), the opacity is greater in the ring. \citet{kra01} also studied the
silicate absorption feature at 10\,$\mu$m, and found a patchy distribution, not 
necessarily correlated to the point sources, the temperature, or the opacity. So
far, dust properties may significantly change in areas as small as a few arcsec, 
Although this result encourages one of our alternatives (variation in grain 
composition), the distribution of HCO and SiO is still an open subject because
the other proposed explanations (different desorption yield and low-velocity shocks)
may be working in this case.

To summarize, our SiO and HCO observations show that these compounds 
are enhanced in the outskirts of the UC \ion{H}{2} regions ($A_{\mathrm V}>5$\,mag). The 
enhancement in the abundance of these species may be related to the interaction of 
the UV photons with grains, instead of the direct effect of the UV photons on the 
molecular gas chemistry. The interaction of the UV photons with grains may 
modify the chemistry of large column densities of the molecular gas, and may
show an evidence of the advancing PDR into the molecular cloud.

\section{On the kinematics of the dense gas layer}

In the UC \ion{H}{2} regions, the high densities of the ionized gas lead to enormous 
internal pressures ($n\,T \sim 10^9$ cm$^{-3}$ K). Since the sources are
very small, their expansion timescales should be as short as a few hundred
years \citep{dre81}. Surprisingly, the number of UC \ion{H}{2} regions is 
greater than the expected from the massive star formation rate \citep{woo89},
which is known as the ``lifetime problem'' of the UC \ion{H}{2} regions. 
A deep knowledge of the morphology and physical conditions of the dense PDR 
around the UC \ion{H}{2} region in \objectname{Mon R2}, would shed some light on this problem.
We can explore one of the proposed explanations, in which the extremely dense 
neutral gas confine the {\sc Hii} region by pressure \citep{dep95,ake96}; our 
data show the existence of a gas layer with $n$(H$_2$)$>4\,10^6$ cm$^{-3}$. 
Hence, if we assume a kinetic temperature of 100\,K, a hydrogen density 
$n$(\htwo) $\sim10^7$ cm$^{-3}$ would be enough to pressure-confine the 
UC \ion{H}{2} region.  

The velocity range associated to C2 is undoubtedly different from that linked 
to the parent molecular cloud, and hence it should have a particular kinematics. 
By means of the PDR chemistry, we have a good chance of further studying 
the kinematics of the molecular gas confining the UC \ion{H}{2} region, because some
chemical species may avoid the confusion of the foreground molecular cloud.
In Fig.~8, we show the position-velocity (P-V) diagram of the 
C$_3$H$_2$ 2$\rightarrow$1 (Fig.~8a) and 6$\rightarrow$5 lines (Fig.~8b) 
across the observed strip. As we previously discussed, most of the C$_3$H$_2$ 
2$\rightarrow$1 line traces C1, whereas the 6$\rightarrow$5 emission is dominated
by C2. The P-V diagram of the C$_3$H$_2$ 6$\rightarrow$5 line is 
preferably located toward the \ion{H}{2} region (center of the map), rather 
than extended along the whole strip. The curves superimposed in Fig.~8 were 
symmetrically traced around C1, and roughly follow the \cthd\ 6$\rightarrow$5
maxima. As the extreme velocities of the curves are $\approx 7$ and 12 \kms, this 
pattern is compatible with a picture where the PDR is expanding at a velocity 
between 2 and 3 \kms. It is worth mentioning that such a low expansion velocity 
was recently observed in an UC \ion{H}{2} region in W48, by observations of carbon 
recombination lines \citep{ros05}.

%% ======== C O N C L U S I O N S ========================================

\section{Conclusions}
We have carried out a molecular survey towards the UC \ion{H}{2} region 
\objectname{Mon R2} to have a deeper insight into the chemistry associated to 
this kind of objects. Our results show the existence of two groups of molecules. 
The first group (HUV) is formed by molecules that present large abundances in the 
surface layers of PDRs ($A_{\mathrm V}<5$\,mag). The species CO$^+$, HOC$^+$, 
C$_2$H, and C$_3$H$_2$ belong to this group. Gas-phase PDR models can successfully 
account for the behavior of CO$^+$, HOC$^+$, and C$_2$H. In the case of 
C$_3$H$_2$, however, PDR models fall short of explaining the observed abundances. 
New C$_3$H$_2$ formation mechanisms linked to the photodestruction of PAHs have 
been proposed to account for this difference.

HCO and SiO belong to another group of molecules (LUV). Their abundances are 
enhanced at the edges of the PDR ($A_{\mathrm V}>5$\,mag), which cannot be 
explained by current gas-phase PDR models. Photodesorption from grain mantles 
appears like a probable mechanism capable of explaining the observed abundances. 
A revision of gas-phase chemical networks, as well as a model able to include 
photodesorption in PDRs, may significantly improve the knowledge of LUV species.
Other molecules, like NH$_3$, whose chemistry is related with 
the grains also present large abundances at the edges of the \ion{H}{2} regions 
\citep{lar03}. This issue may show that the interactions of the UV radiation with the 
grains could produce important effects on the chemistry of the molecular gas, 
even at local visual extinction as deep as 10\,mag.

Finally, we have used PDR chemistry to determine the physical conditions and 
kinematics of the layer of gas surrounding the UC \ion{H}{2} region. Using a LVG code 
to fit our C$_3$H$_2$ observations, we have derived hydrogen densities 
$>4\,10^6$ cm$^{-3}$ for this layer. The proposed scenario of a pressure-confined 
\ion{H}{2} region requires $n$(\htwo) $\sim 10^7$ cm$^{-3}$, perfectly compatible with 
this result. In addition, the P-V diagram of the C$_3$H$_2$ lines might show that 
while the gas layers at a visual extinction $<5$\,mag are expanding at a velocity 
of $\sim$ 2--3 km s$^{-1}$, the back layers of the PDR are moving at the velocities 
of the foreground molecular cloud. This kinematics supports the 
interpretation that the UC \ion{H}{2} region could be confined by a dense 
($n$(\htwo) $>10^6$ cm$^{-3}$) and thin ($<4$ 10$^{21}$ cm$^{-2}$) layer of 
neutral gas.

On the other hand, the search for other LUV molecules looks as an exciting 
observational task which may lead to determine the chemical processes carrying out 
in the more shielded part of PDRs. Moreover, further observations and 
multi-transitional studies of the HUV molecules are needed to determine 
accurately the physical conditions and dynamics of this neutral layer.

\acknowledgements
We are grateful to the technical staff in Pico de Veleta for their professional 
support during the observations. We specially wish to thank the anonymous referee,
who greatly improved the paper by sharing with us his/her viewpoints about the
PDR chemistry.
This paper has been partially funded by the Spanish 
MCyT under projects DGES/AYA2000-927, ESP2001-4519-PE, ESP2002-01693, 
and AYA2003-06473.

%% To help institutions obtain information on the effectiveness of their
%% telescopes, the AAS Journals has created a group of keywords for telescope
%% facilities. A common set of keywords will make these types of searches
%% significantly easier and more accurate. In addition, they will also be
%% useful in linking papers together which utilize the same telescopes
%% within the framework of the National Virtual Observatory.
%% See the AASTeX Web site at http://www.journals.uchicago.edu/AAS/AASTeX
%% for information on obtaining the facility keywords.

%% After the acknowledgments section, use the following syntax and the
%% \facility{} macro to list the keywords of facilities used in the research
%% for the paper.  Each keyword will be checked against the master list during
%% copy editing.  Individual instruments can be provided in parentheses,
%% after the keyword, but they will not be verified.

Facilities: \facility{IRAM(MRT-30m)}.

%% The reference list follows the main body and any appendices.
%% Use LaTeX's thebibliography environment to mark up your reference list.
%% Note \begin{thebibliography} is followed by an empty set of
%% curly braces.  If you forget this, LaTeX will generate the error
%% "Perhaps a missing \item?".
%%
%% thebibliography produces citations in the text using \bibitem-\cite
%% cross-referencing. Each reference is preceded by a
%% \bibitem command that defines in curly braces the KEY that corresponds
%% to the KEY in the \cite commands (see the first section above).
%% Make sure that you provide a unique KEY for every \bibitem or else the
%% paper will not LaTeX. The square brackets should contain
%% the citation text that LaTeX will insert in
%% place of the \cite commands.

%% We have used macros to produce journal name abbreviations.
%% AASTeX provides a number of these for the more frequently-cited journals.
%% See the Author Guide for a list of them.

%% Note that the style of the \bibitem labels (in []) is slightly
%% different from previous examples.  The natbib system solves a host
%% of citation expression problems, but it is necessary to clearly
%% delimit the year from the author name used in the citation.
%% See the natbib documentation for more details and options.

\clearpage

%% Use the figure environment and \plotone or \plottwo to include
%% figures and captions in your electronic submission.
%% To embed the sample graphics in
%% the file, uncomment the \plotone, \plottwo, and
%% \includegraphics commands
%%
%% If you need a layout that cannot be achieved with \plotone or
%% \plottwo, you can invoke the graphicx package directly with the
%% \includegraphics command or use \plotfiddle. For more information,
%% please see the tutorial on "Using Electronic Art with AASTeX" in the
%% documentation section at the AASTeX Web site,
%% http://www.journals.uchicago.edu/AAS/AASTeX.
%%
%% The examples below also include sample markup for submission of
%% supplemental electronic materials. As always, be sure to check
%% the instructions to authors for the journal you are submitting to
%% for specific submissions guidelines as they vary from
%% journal to journal.

%%------- F I G U R E   1 -----------------------

\begin{figure}[!ht]
\epsscale{.42}
\plotone{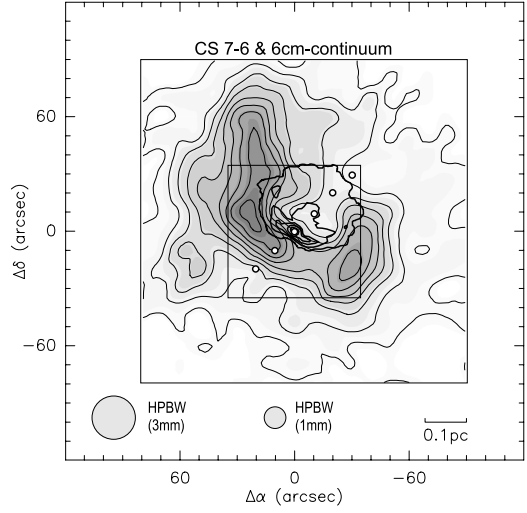}
\caption{
Location of the positions observed in this paper. Greyscale
represent the CS 7$\rightarrow$6 line emission map from \citet{cho00}. The 
circles across the symmetry axis of the UC {\sc Hii} region indicate the positions
observed in the \cthd\ and \cdh\ lines. The small square, $72''\times72''$ in 
size, traces the mapped area in HC$_3$N, H$^{13}$CO$^+$, SiO, and HCO.
}
\end{figure}
%%-----------------------------------------------

%%------- F I G U R E   2 -----------------------
\clearpage
\begin{figure}
\epsscale{1.}
\plotone{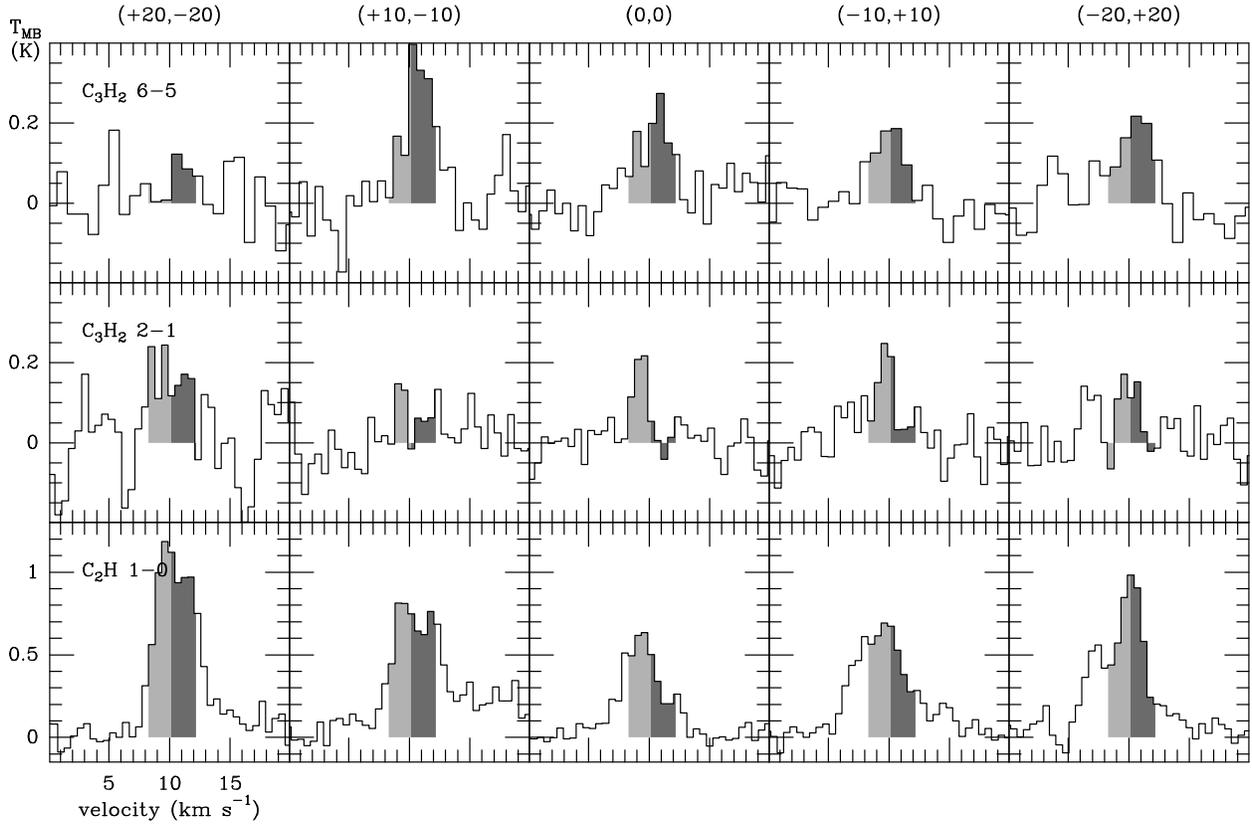}
\caption{
A sample of \cthd\ 6$\rightarrow$5 (\textbf{a}), \cthd\ 2$\rightarrow$1 
(\textbf{b}), and C$_2$H 1$\rightarrow$0 (\textbf{c}) spectra along the strip. 
The positions are indicated at the top part of the figure. The shadowed areas 
indicate the approximate velocity extension of the two kinematic component 
present in the region.
}
\end{figure}
%%----------------------------------------------

%%------- F I G U R E   3 -----------------------
\clearpage
\begin{figure}
\epsscale{1.}
\epsscale{0.4}
\plotone{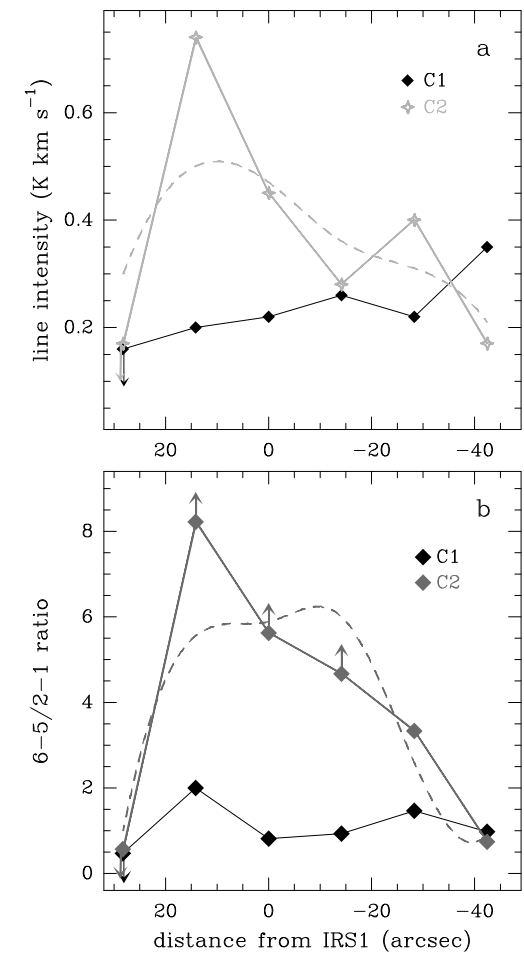}
\caption{
\cthd\ observed parameters along the strip, for each of the velocity components
defined in the text. (\textbf{a}) \cthd\ 6$\rightarrow$5 line intensity. (\textbf{b}) 
\cthd\ 6$\rightarrow$5 to 2$\rightarrow$1 line ratio. Note the uniformity of the C1 
component and the separation of both components in Fig.~b. The dashed curves are the
corresponding to the C2 component, after smoothing the 6$\rightarrow$5 line to the 
resolution of the 2$\rightarrow$1 line. Angular distances are referred to Mon R2 
IRS\,1, increasing toward the southeast.
}
\end{figure}
%%-----------------------------------------------

%%------- F I G U R E   4 -----------------------
\clearpage
\begin{figure}[!ht]
\epsscale{1.}
\plotone{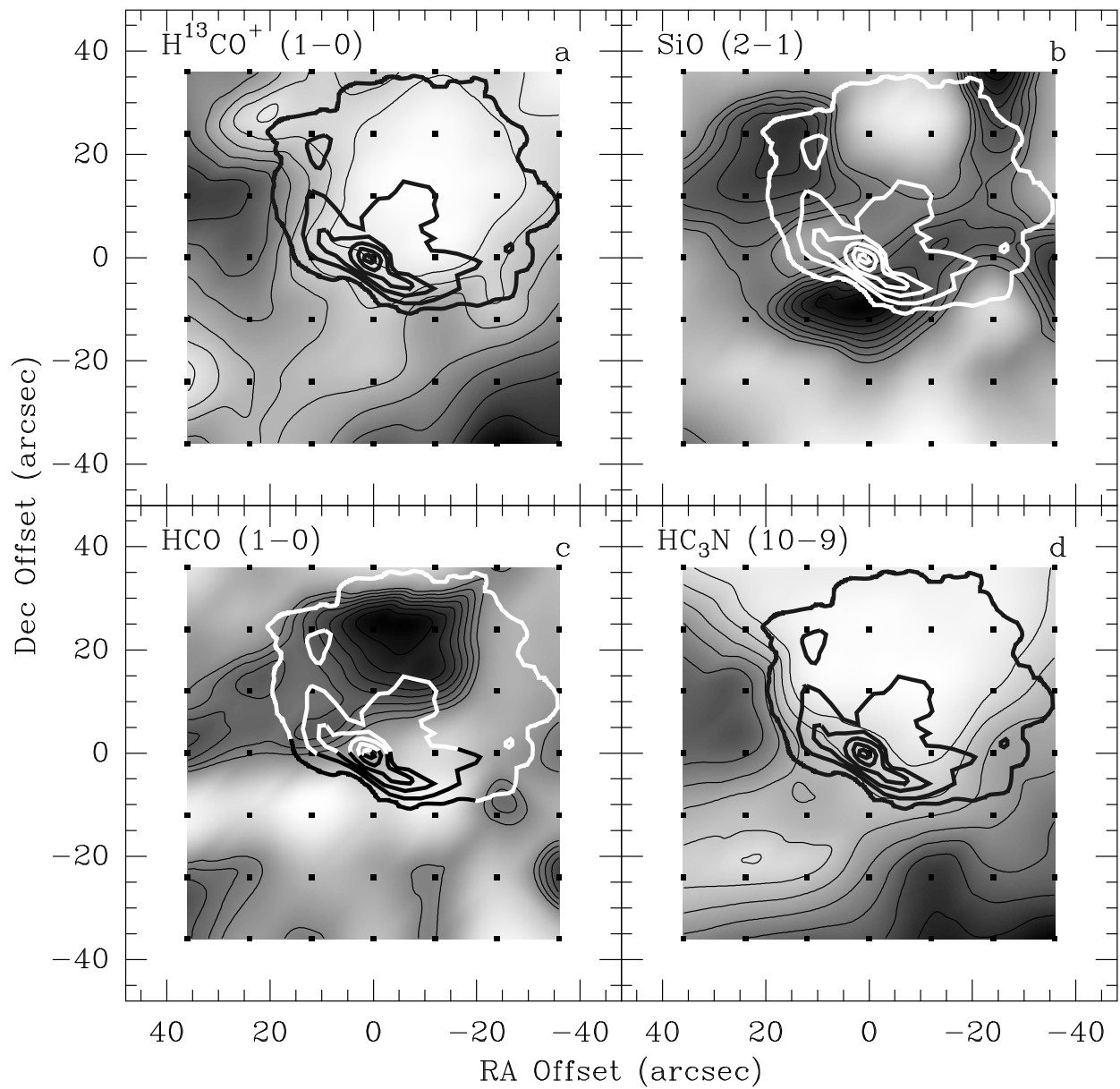}
\caption{
Emission maps around the UC {\sc Hii} region Mon R2. The mapped rotational
lines are indicated in the top left corner of each map. Contour levels are
30\,\% to 90\,\% of the map peak, in steps of 10\,\%. The map peaks are 
2.265, 0.243, 0.244, and 3.651 K\,\kms\ for maps {\textbf a} to {\textbf d}, 
respectively. Superimposed are the 6 cm-continuum emission from \citet{woo89}.
}
\end{figure}
%%-----------------------------------------------

%%------- F I G U R E   5 -----------------------
\clearpage
\begin{figure}
\epsscale{0.6}
\plotone{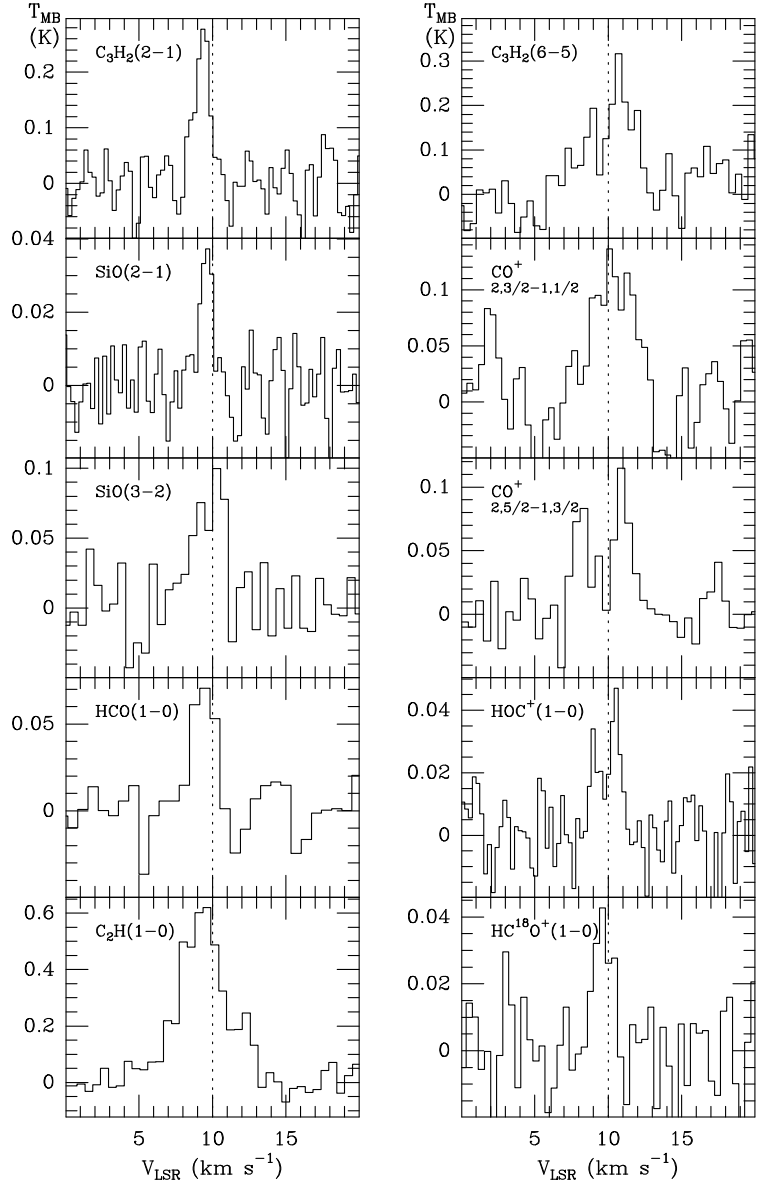}
\caption{
Spectra toward the position (0\arcsec, 0\arcsec). The rotational transition observed is
indicated. Temperature scale is $T_{\mathrm MB}$, in K. There are two types of 
emission depending on the velocity range, roughly in agreement with the velocity 
components defined in Sect.~3. The emission below 10 \kms\ is associated to the 
molecular envelope (component 1), while the emission above 10 \kms\ is linked to
the ultracompact \ion{H}{2} region (component 2).
}
\end{figure}
%%-----------------------------------------------

%%------- F I G U R E   6 -----------------------
\clearpage
\begin{figure}
\epsscale{1.}
\plotone{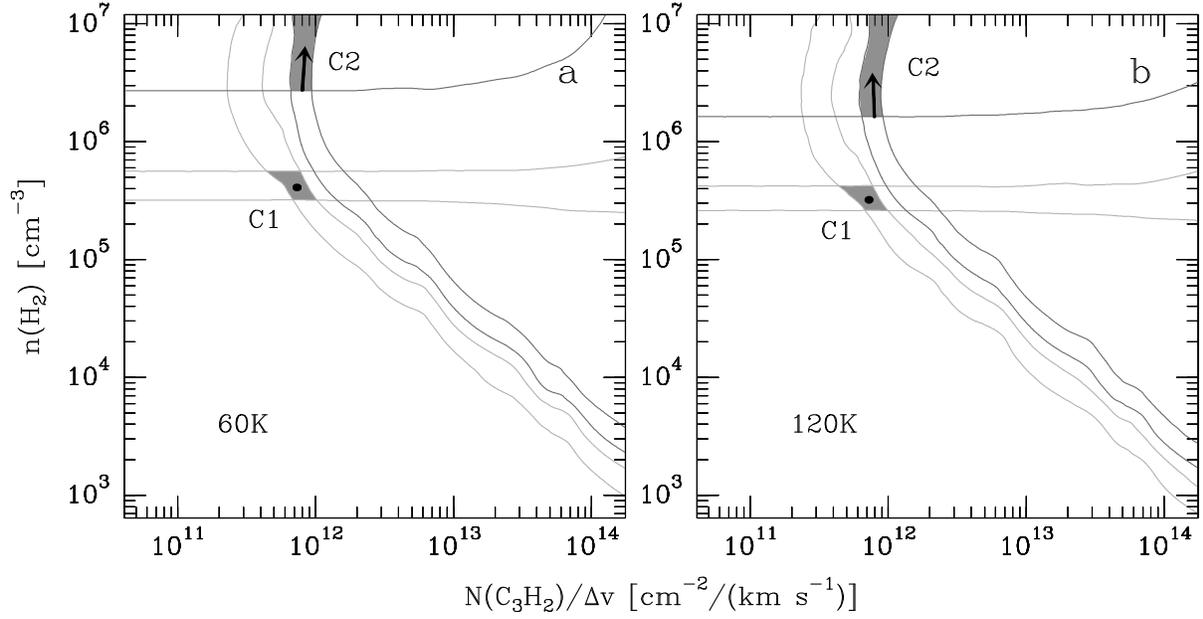}
\caption{
LVG results at the ($0"$,$0"$) position, for kinetic temperatures of 
(\textbf{a}) 60\,K and (\textbf{b}) 120\,K. For each component, it is plotted 
the observed values of the C$_3$H$_2$ 6$\rightarrow$5 line intensity (vertical/diagonal 
lines), as well as the C$_3$H$_2$ [6$\rightarrow$5]/[2$\rightarrow$1] line ratio (horizontal lines), as 
functions of the C$_3$H$_2$ column density and the H$_2$ density. Shadowed areas take 
into account the observational errors. Although both components have comparable 
column densities, the C1 density is lower than that corresponding to C2 at least by
a factor of four.
}
\end{figure}
%%-----------------------------------------------

%%------- F I G U R E   7 -----------------------
\clearpage
\begin{figure}
\epsscale{0.5}
\plotone{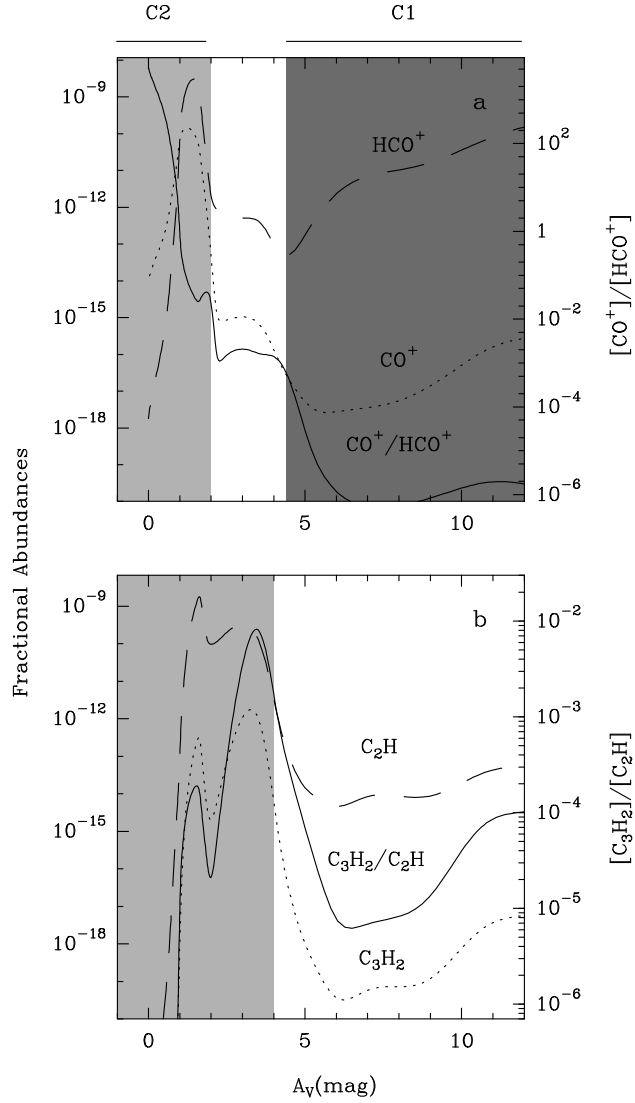}
\caption{Model predictions for the abundances of various species derived from 
updated Le Bourlot et al.'s model. The calculations have been carried out for 
$n$(\htwo)=2 10$^6$ cm$^{-3}$ and G$_0$=5 10$^5$ in units of the Habing field. 
(\textbf{a}) CO$^+$ and HCO$^+$ fractional abundances. CO$^+$ is significantly 
abundant at visual extinctions around 2\,mag, indicated by the light gray area,
which roughly corresponds to the values measured in C2. On the other hand, the 
dark gray area agrees with the upper limits of the [CO$^+$/HCO$^+$] ratio 
measured in C1. (\textbf{b}) \cthd\ and \cdh\ fractional abundances, as a 
function of the visual extinction. The [\cthd]/[\cdh] ratio peaks between 2 
and 5\,mag (gray areas).
}
\end{figure}
%%-----------------------------------------------

%%------- F I G U R E   8 -----------------------
\clearpage
\begin{figure}
\epsscale{0.5}
\plotone{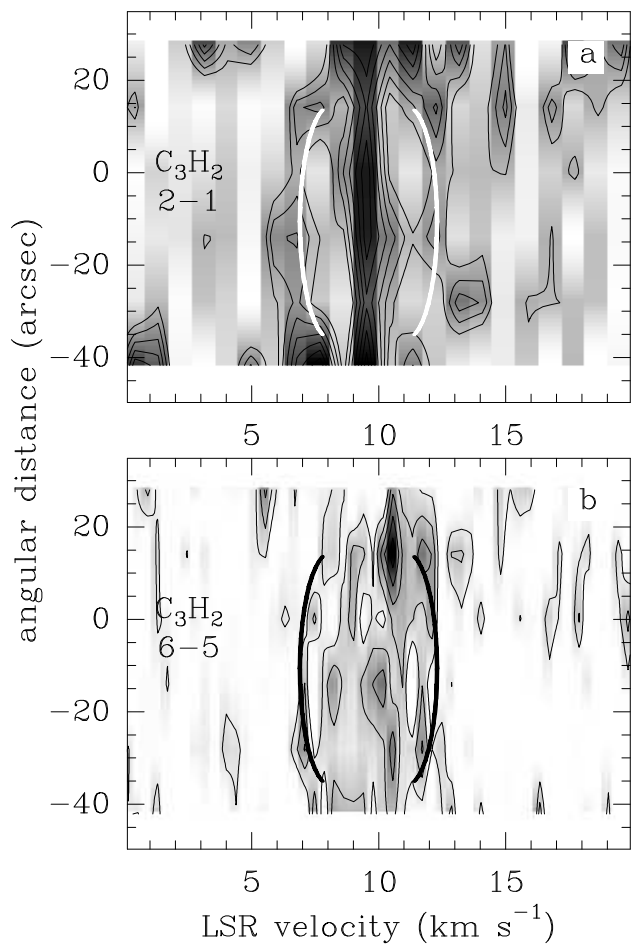}
\caption{
Position-velocity diagram of (\textbf{a}) \cthd\ 2$\rightarrow$1 and 
(\textbf{b}) \cthd\ 6$\rightarrow$5 lines across the strip. Angular distances 
are referred to Mon R2 IRS\,1, increasing toward the southeast. Most of the
\cthd\ 2$\rightarrow$1 emission arises from the C1 component, between 9 and 10
\kms. The curves roughly follow the maxima at high velocities, which includes
the C2 component. This pattern is compatible with an expansion of 2--3 \kms
}
\end{figure}

\clearpage

%%--------  T A B L E   1 ------------------------
\begin{deluxetable}{lcrcc}
\tablecaption{Observing frequencies and telescope parameters}
\tablecolumns{5}
\tablewidth{0pc}

\tablehead{
\colhead{Molecule} & \colhead{Transition} & 
\colhead{Freq (GHz)} &
\colhead{beam} & \colhead{$\eta_{MB}$}
}

\startdata
C$_2$H & $N=1-0, J=3/2-1/2$ \\
& $F=1\rightarrow1$  &  87.284156 & 28.5\arcsec  & 0.77  \\
& $F=2\rightarrow1$  &  87.316925 & 28.5\arcsec  & 0.77  \\
& $F=1\rightarrow0$  &  87.328624 & 28.5\arcsec  & 0.77  \\
\\
C$_3$H$_2$ & $J(K_a,K_c) =$ \\
& 2(1,2) $\rightarrow$ 1(0,1)  &  85.338893 & 29\arcsec  & 0.77  \\
& 3(2,2) $\rightarrow$ 2(1,1)  & 155.518295 & 17\arcsec  & 0.65  \\
& 6(0,6) $\rightarrow$ 5(1,5)  & 217.822148 & 11.5\arcsec  & 0.52 \\
& 6(1,6) $\rightarrow$ 5(0,5)  & 217.822148 & 11.5\arcsec  & 0.52 \\
& 5(1,4) $\rightarrow$ 4(2,3)  & 217.940046 & 11.5\arcsec  & 0.52 \\
\\
HC$_3$N & $J=10 \rightarrow 9$             &  90.9789933 & 27\arcsec & 0.76 \\
\\
HCO & $N_{K-K+}=1_{01}-0_{00}$ \\
& $J=3/2-1/2, F=2\rightarrow 1$              &  86.6708200 & 28.5\arcsec & 0.77   \\
\\
SiO & $J=2\rightarrow 1$             &  86.8468910 & 28.5\arcsec & 0.77 \\
SiO & $J=3\rightarrow 2$              & 130.268702 & 19\arcsec & 0.70 \\
\\
H$^{13}$CO$^+$ & $J=1\rightarrow 0$ & 86.7543300 & 28.5\arcsec & 0.77 \\
\enddata

\end{deluxetable}

%\clearpage

%=========  T A B L E   2 ===========================================
\begin{deluxetable}{clrrr}
\tablecaption{One-component Gaussian fits to the \cthd\ and \cdh\ spectral lines}
\tablecolumns{5}
\tablewidth{0pc}

\tablehead{
\colhead{Offset} & \colhead{Line} &
\colhead{Area} &
\colhead{V$_{\rm LSR}$} &
\colhead{$\Delta$V$_{1/2}$} \\

\colhead{arcsec} & &
\colhead{K km s$^{-1}$ } &
\colhead{km s$^{-1}$} &
\colhead{km s$^{-1}$}
}

\startdata
(20, -20)  & \cthd\ $6\rightarrow5$\tablenotemark{a} & $<$ 0.13 & \nodata & \nodata \\
(10, -10)  & \cthd\ $6\rightarrow5$\tablenotemark{a} & 1.10 (0.10) & 10.9 (0.1) & 2.8 (0.3) \\
(0, 0)     & \cthd\ $6\rightarrow5$\tablenotemark{a} & 0.82 (0.11) & 10.4 (0.2) & 3.7 (0.6) \\
(-10, +10) & \cthd\ $6\rightarrow5$\tablenotemark{a} & 0.76 (0.10) &  9.8 (0.2) & 3.6 (0.6) \\
(-20, +20) & \cthd\ $6\rightarrow5$\tablenotemark{a} & 0.76 (0.15) & 10.5 (0.3) & 3.3 (0.8) \\
(-30, +30) & \cthd\ $6\rightarrow5$\tablenotemark{a} & 0.70 (0.14) &  9.5 (0.3) & 2.7 (0.5) \\
\\
(20, -20)  & \cthd\ $5(1,4)\rightarrow4(2,3)$ & $<$ 0.07 & \nodata & \nodata \\
(10, -10)  & \cthd\ $5(1,4)\rightarrow4(2,3)$ & 0.76 (0.09) & 10.6 (0.3) & 3.4 (0.5) \\
(0, 0)     & \cthd\ $5(1,4)\rightarrow4(2,3)$ & 0.56 (0.07) &  8.9 (0.3) & 4.2 (0.6) \\
(-10, +10) & \cthd\ $5(1,4)\rightarrow4(2,3)$ & 0.47 (0.12) & 10.6 (0.6) & 4.4 (1.2) \\
(-20, +20) & \cthd\ $5(1,4)\rightarrow4(2,3)$ & $<$ 0.06 & \nodata & \nodata \\
(-30, +30) & \cthd\ $5(1,4)\rightarrow4(2,3)$ & $<$ 0.06 & \nodata & \nodata \\
\\
(20, -20)  & \cthd\ $3(2,2)\rightarrow2(1,1)$ & $<$ 0.11 & \nodata & \nodata \\
(10, -10)  & \cthd\ $3(2,2)\rightarrow2(1,1)$ & 0.18 (0.04) & 11.6 (0.1) & 0.5 (0.2) \\
(0, 0)     & \cthd\ $3(2,2)\rightarrow2(1,1)$ & 0.38 (0.09) &  9.8 (0.7) & 5.6 (1.3) \\
(-10, +10) & \cthd\ $3(2,2)\rightarrow2(1,1)$ & $<$ 0.09 & \nodata & \nodata \\
(-20, +20) & \cthd\ $3(2,2)\rightarrow2(1,1)$ & $<$ 0.11 & \nodata & \nodata \\
(-30, +30) & \cthd\ $3(2,2)\rightarrow2(1,1)$ & $<$ 0.15 & \nodata & \nodata \\
\\
(20, -20)  & \cthd\ $2(1,2)\rightarrow1(0,1)$ & 0.62 (0.20) &  9.9 (0.6) & 3.3 (1.1) \\
(10, -10)  & \cthd\ $2(1,2)\rightarrow1(0,1)$ & $<$ 0.11 & \nodata & \nodata \\
(0, 0)     & \cthd\ $2(1,2)\rightarrow1(0,1)$ & 0.29 (0.04) &  9.3 (0.1) & 1.1 (0.2) \\
(-10, +10) & \cthd\ $2(1,2)\rightarrow1(0,1)$ & 0.41 (0.09) &  9.7 (0.2) & 1.6 (0.6) \\
(-20, +20) & \cthd\ $2(1,2)\rightarrow1(0,1)$ & 0.25 (0.09) &  9.8 (0.3) & 1.8 (0.7) \\
(-30, +30) & \cthd\ $2(1,2)\rightarrow1(0,1)$ & 1.09 (0.17) &  8.4 (0.4) & 4.6 (0.7) \\
\\
(20, -20)  & \cdh\ $1\rightarrow0$ & 4.77 (0.18) & 10.5 (0.1) & 3.9 (0.2) \\
(10, -10)  & \cdh\ $1\rightarrow0$ & 3.49 (0.03) & 10.7 (0.2) & 4.7 (0.3) \\
(0, 0)     & \cdh\ $1\rightarrow0$ & 2.32 (0.12) &  9.3 (0.1) & 3.8 (0.3) \\
(-10, +10) & \cdh\ $1\rightarrow0$ & 3.67 (0.10) &  9.3 (0.1) & 5.3 (0.2) \\
(-20, +20) & \cdh\ $1\rightarrow0$ & 3.57 (0.18) &  9.7 (0.1) & 4.5 (0.3) \\
(-30, +30) & \cdh\ $1\rightarrow0$ & 1.97 (0.23) & 10.3 (0.1) & 2.7 (0.5) \\
\enddata

\tablecomments{Number in parenthesis are 3$\sigma$ errors in each parameter.}

\tablenotetext{a}{Parameters correspond to the transitions $6(1,6)\rightarrow5(0,5)$ and 
         $6(0,6)\rightarrow5(1,5)$ together, which are blended at the same frequency.}

\end{deluxetable}
%%--------------------------------------------------

\clearpage

%=========  T A B L E   3 ===========================================
\begin{deluxetable}{crrcrrcrr}
\tablecaption{LVG results for \cthd\ and C$_2$H}
\tablecolumns{9}
\tablewidth{0pc}

\tablehead{
\colhead{}
&
\multicolumn {2}{c}{Component 1} &&
\multicolumn {2}{c}{Component 2} && \\

\noalign{\smallskip}
\cline{2-3}
\cline{5-6}
\cline{8-9} \\

\colhead{Offset} & 
\colhead{$n$(H$_2$)} &
\colhead{$N$(C$_3$H$_2$)} &&
\colhead{$n$(H$_2$)} &
\colhead{$N$(C$_3$H$_2$)} &&
\colhead{$N$(C$_2$H)} &
\colhead{$\frac{[{\mathrm {C_3H_2}}]}{[{\mathrm {C_2H}}]}$}\tablenotemark{a} \\

\colhead{(\arcsec,\arcsec)} & 
\colhead{cm$^{-3}$} & 
\colhead{10$^{12}$ cm$^{-2}$} &&
\colhead{cm$^{-3}$} & 
\colhead{10$^{12}$ cm$^{-2}$} &&
\colhead{10$^{14}$ cm$^{-2}$} & 
\colhead{}
}

\startdata
(+20, -20) & $<$2.4 10$^5$ & $\la$2.2 && $<$4.6 10$^5$ & $\la$1.8 && 4.2 & 0.010 \\
(+10, -10) &    8.4 10$^5$ &      1.1 && $>$3.5 10$^6$ & 4.0      && 3.0 & 0.017 \\
(0, 0)     &    4.5 10$^5$ &      1.8 && $>$4.5 10$^6$ & 2.3      && 2.0 & 0.021 \\
(-10, +10) &    4.6 10$^5$ &      2.2 && $>$4.7 10$^6$ & 1.4      && 3.2 & 0.011 \\
(-20, +20) &    6.4 10$^5$ &      1.6 &&    1.1 10$^6$ & 2.4      && 3.1 & 0.013 \\
(-30, +30) &    4.5 10$^5$ &      3.1 &&    4.4 10$^5$ & 1.8      && 1.7 & 0.029 \\
\enddata

\tablecomments{Assumed a kinetic temperature of 50\,K. C$_2$H was not separated 
               by components.}

\tablenotetext{a}{The $\frac{[{\mathrm {C_3H_2}}]}{[{\mathrm {C_2H}}]}$ abundance ratio involves the 
whole emission from the two molecules.}

\end{deluxetable}

\clearpage
%=========  T A B L E   4 ===========================================
\begin{deluxetable}{lcrrrr}
\tablecolumns{6}
\tablecaption{Column densities and relative abundances}
\tablewidth{0pc}

\tablehead{
\colhead{Molecule/ratio} &
\colhead{unit} &
\colhead{C1(10,-10)} &
\colhead{C2(10,-10)} &
\colhead{C1(0,0)} &
\colhead{C2(0,0)} 
}

\startdata
\cthd\  & $10^{12}$ cm$^{-2}$ & 1.1    & 4.0    & 1.8    & 2.3 \\
C$_2$H  & $10^{14}$ cm$^{-2}$ & 1.4    & 1.5    & 1.2    & 0.8 \\
CO$^+$  & $10^{10}$ cm$^{-2}$ & $<4.8$ & $<5.2$ & $<4.8$ & 31.1 \\
HOC$^+$ & $10^{10}$ cm$^{-2}$ & $<2.6$ & 9.9    & 8.8    & 9.6 \\
HCO     & $10^{12}$ cm$^{-2}$ & $<3.6$ & $<3.9$ & 3.7    & $<1.2$ \\
H$^{13}$CO$^+$ 
        & $10^{11}$ cm$^{-2}$ & 9.3    & 9.5    & 5.9    & 5.8 \\
HC$_3$N & $10^{12}$ cm$^{-2}$ & 2.0    & 5.2    & 1.1    & 4.0 \\
SiO     & $10^{10}$ cm$^{-2}$ & 34     & $<22$  & 11     & $<1.6$ \\
HC$^{18}$O$^+$
        & $10^{10}$ cm$^{-2}$ & 10.9   & 14.2   & 11.3   & 6.6 \\
\\
HCO$^+$/HOC$^+$ &            & $>2700$ & 930    & 830    & 450 \\
CO$^+$/HCO$^+$  & $10^{-4}$  & $<7$    & $<6$   & $<7$   & 72 \\
%CO$^+$/HOC$^+$ &            & ...     & $<0.5$ & $<0.6$ & 3.2 \\
\\

\cthd/C$_2$H &  $10^{-3}$    & 8       & 26     & 15     & 30 \\
\cthd/HC$_3$N &              & 0.6     & 0.8    & 1.6    & 0.6 \\
HCO/H$^{13}$CO$^+$ &         & $<3.9$  & $<4.1$ & 6.3    & $<2.1$ \\
SiO/H$^{13}$CO$^+$ &         & 0.37    & $<0.23$& 0.19   & $<0.03$ \\
%\cthd/HCO$^+$     &         & 0.02    & 0.09   & 0.03   & 0.12 \\
%C$_2$H/HCO$^+$    &         & 1.6     & 1.4    & 1.0    & 1.0 \\
\enddata

\end{deluxetable}
%%--------------------------------------------------

\clearpage
%=========  T A B L E   5 ===========================================

%\begin{deluxetable}{cr@{\qquad}r@{\qquad}r@{\qquad}r@{\qquad}r@{\quad}r@{\quad}r@{\quad}l}
\begin{deluxetable}{crrrrrrrl}
\rotate
\tablecolumns{9}
\tablewidth{0pc}
\tabletypesize{\small}
\tablecaption{Column densities and abundances at selected positions}

\tablehead{
\colhead{Position} & 
\colhead{$N$(HCO)} &
\colhead{$N$(H$^{13}$CO$^+$)} &
\colhead{$N$(HC$_3$N)} &
\colhead{$N$(SiO)} &
\colhead{$\frac{[{\rm HCO}]}{[{\rm H^{13}CO^+}]}$} &
\colhead{$\frac{[{\rm HCO}]}{[{\rm HC_3N}]}$} &
\colhead{$\frac{[{\rm HCO}]}{[{\rm SiO}]}$} &
\colhead{Comment} \\

\colhead{(\arcsec,\arcsec)} & 
\colhead{$10^{12}$ cm$^{-2}$} & 
\colhead{$10^{12}$ cm$^{-2}$} & 
\colhead{$10^{12}$ cm$^{-2}$} & 
\colhead{$10^{11}$ cm$^{-2}$} &
\colhead{} & 
\colhead{} & 
\colhead{} & 
\colhead{} 

}

\startdata

(+12, +24) & 15.8 & 1.7 & 6.4 & $<2.2$ &  9.3 & 2.5 & $>720$     & HCO peak \\
(0, +24)   & 14.1 & 1.0 & 4.1 & $<2.2$ & 14.7 & 3.4 & $>640$     & HCO peak \\
(0, 0)   & $<2.5$ & 1.6 & 11.9 & 0.8 & $<1.6$ & $<0.2$ & $<310$  & center \\
(+12, -12) & $<3.7$ & 1.7 & 11.1 & 4.3 & $<2.2$ & $<0.3$ & $<90$ & SiO peak \\
%(+12, 0)   &  6.9 & 1.2 & 10.0 & 2.6 & 5.8 & 0.7 & 270 \\
(+36, +12) &  4.0 & 2.8 & 18.9 & $<2.0$ & 1.4 & 0.2 & $>200$ & Dense envelope \\
(-12, -36) &  9.6 & 1.8 & 25.6 & $<1.6$ & 5.3 & 0.4 & $>600$ & Dense envelope \\
(0, -12) & $<3.1$ & 1.3 & 12.0 & 6.0 & $<2.4$ & $<0.3$ & $<50$   & SiO peak \\

\enddata

\end{deluxetable}
%%--------------------------------------------------

%% Tables may also be prepared as separate files. See the accompanying
%% sample file table.tex for an example of an external table file.
%% To include an external file in your main document, use the \input
%% command. Uncomment the line below to include table.tex in this
%% sample file. (Note that you will need to comment out the \documentclass,
%% \begin{document}, and \end{document} commands from table.tex if you want
%% to include it in this document.)

%% \input{table}

%% The following command ends your manuscript. LaTeX will ignore any text
%% that appears after it.

\end{document}